# The effectiveness of using Google Maps Location History data to detect joint activities in social networks

Giancarlos Parady
gtroncoso@ut.t.u-tokyo.ac.jp
Department of Urban Engineering, The University of Tokyo

Keita Suzuki
IBM Japan

Yuki Oyama
Shibaura Institute of Technology

Makoto Chikaraishi
Hiroshima University

## Abstract
This study evaluates the effectiveness of using Google Maps Location History data to identify joint activities in social networks. To do so, an experiment was conducted where participants were asked to execute daily schedules designed to simulate daily travel incorporating joint activities. For Android devices, detection rates for 4-person group activities ranged from 22% under the strictest spatiotemporal accuracy criteria to 60% under less strict yet still operational criteria. The performance of iPhones was markedly worse than Android devices, irrespective of accuracy criteria. In addition, logit models were estimated to evaluate factors affecting activity detection given different spatiotemporal accuracy thresholds. In terms of effect magnitudes, non-trivial effects on joint activity detection probability were found for floor area ratio (FAR) at location, activity duration, Android device ratio, device model ratio, whether the destination was an open space or not, and group size.
   Although current activity detection rates are not ideal, these levels must be weighed against the potential of observing travel behavior over long periods of time, and that Google Maps Location History data could potentially be used in conjunction with other data-gathering methodologies to compensate for some of its limitations.

# 1. Introduction

Our travel patterns are interdependent with the travel patterns of our families, friends, colleagues, and other members of our social networks. We coordinate with the people in our network to conduct joint activities for support, leisure, and other purposes (Carrasco and Miller, 2006), making social activities account for a significant share of trips and one of the fastest-growing segments of travel (Axhausen, 2005). For example, joint trips account for 40% to 60% of all out-of-home activities in Japan (Qian et al., 2019). Furthermore, social activities (leisure in particular) have lower frequencies than work or maintenance trips, and high degrees of spatiotemporal variability, which makes them (i) very hard to capture in traditional studies using one-weekday travel diaries that do not explicitly account for social network characteristics and joint activities, and (ii) very hard to predict. As such, social activities remain poorly explained in traditional travel behavior models, and lack of empirical data remains a key obstacle towards a more adequate incorporation of joint activities and social interactions in transportation studies. To date, most studies on joint activities have used an agent-based simulation framework (Arentze and Timmermans, 2008; Ronald, Arentze and Timmermans, 2012), but empirical data is necessary for parameter calibration and model validation.

While in recent years there has been a growing interest in social networks, interactions among network members and its relation to socially motivated travel (Parady, Takami and Harata, 2020, Parady et al, 2021), collecting data on joint activities remains a difficult task given the high cost (in terms of survey execution and response burden) of gathering detailed social networks data in addition to travel behavior.

For example, a study attempting to collect joint activities would have to (i) collect socio-demographic information on a set of respondents (egos), (ii) use name generators and name interpreters to elicit the members (alters) that compose such networks, (iii) conduct a travel behavior survey including companionship questions. One example is the study conducted by Calastri, dit Sourd and Hess (2018), but even in such an extensive study, the behavior of alters (social network members, from the perspective of the respondent) is only observed when interacting with egos (the respondents), and the rest of the activity patterns of alters remain unknown.

Against this background, this study evaluates the effectiveness of using Google Maps Location History data (hereinafter GLH data) to identify joint activities in social networks. To do so, an experiment was conducted where participants were asked to execute daily schedules designed to simulate daily travel incorporating joint activities.

The use of GLH data is an attractive option for several reasons: (i) most people already have the Google Maps application installed in their smartphones and use it frequently, (ii) due to its passive nature there is effectively no burden to respondents, (iii) users can easily download their location history data, and (iv) location history data can be edited by the user. These reasons raise the possibility of using it as a cost-effective passive travel-diary survey tool. While a few studies have evaluated its potential as an activity-travel data collection tool, no study has evaluated its effectiveness in the context of joint activities in social networks.

The rest of this article is structured as follows: Section 2 reviews the relevant literature on the subject. Section 3 summarizes the experiment's protocol. Section 4 briefly describes the GLH data structure, Section 5 describes the accuracy measures used. Section 6 summarizes the main findings. Section 7 briefly discusses issues related to the execution of an empirical study, while Section 8 wraps up by discussing the key implications of this study and identifying further avenues of research.

## 2. Findings from the literature

Because of its passive nature, GLH data is appealing to researchers on human mobility and transportation. While the literature is rather limited, several studies have evaluated GLH data performance in the past (See Cools et al. (2021) for a discussion on other travel passive data collection methods.)

Sadeghvaziri, Rojas IV and Jin (2016) conducted one of the earliest pilot studies and used GLH data to evaluate travel patterns, particularly focusing on location detection, trip purpose and travel mode, highlighting the potential of this data. More recently Ruktanonchai et al. (2018) evaluated the spatial agreement between GLH data for Android users and GPS systems. They found that the two datasets had an 85% agreement when spatially aggregating the data to 100m grid cells without interpolation. They also found that when using interpolated data, at the 100m x 100m resolution the agreement dropped to 60% but increased to 85% when the resolution was decreased to 500m grid cells.

Similarly, Macarulla Rodriguez et al. (2018) compared GLH against a GPS device with higher accuracy and evaluated its location accuracy on different networks (2G, 3G, Wi-Fi) and for mobile devices with GPS. They also evaluated accuracy on different environments (urban vs. rural) and different travel modes (Bike, car, tram, walking) and while static. They found that GPS yielded the best performance, followed by 3G and 2G, respectively, with Wi-fi having the worst performance.

Both of these studies used a GPS device as ground truth data. However, GPS devices are also subject to measurement error. Cools et al. (2021) addressed this issue by conducting an experiment where subjects executed a predefined schedule and self-report arrival and departure times, thus allowing the researcher to get a-priori knowledge of what GLH data should register. They found that GLH had an overall detection rate (referring to detection of locations independently) of 51% and an overall trip detection rate of 32% (referring to a consecutive sequence of two detected locations.) They also showed that shorter dwell times were more likely to be missed by GLH and that iPhones underperformed against Androids with an average location detection rate of 28% against a 57% for Android devices.

Based on these results, they conclude that GLH is not currently an adequate tool to collect travel diary data. However, we argue that these detection rates, if not ideal, must be weighed against the potential of observing travel behavior over long periods of time. For instance, travel diaries are usually limited to one day and suffer from underreporting of activities (Zhao et al., 2015), in particular shorter ones. Passive methods such as GLH can be used as a complement to other data-gathering methodologies given a proper understanding of their accuracy level. However, to do so, more insight is needed on the extent to which potential factors, such as activity duration, built environment characteristics, and group size, affect accuracy and activity detection rates. As such, in spite of the limitations of GLH data, given the issues described above, the ubiquitouness of the Google Maps app and the low burden imposed on respondents, the performance of GLH data for identifying joint activities in social networks is worth evaluating.

## 3. Experiment protocol

### 3.1. Recruitment, schedule design and execution

Participants were students recruited from the University of Tokyo, located in one of the central wards of the Tokyo Metropolis, and Hiroshima University, located in a suburban town of Hiroshima prefecture. These two locations were chosen in order to capture a wider variation in built environment characteristics. For each experiment day, 4 participants were asked to execute a schedule designed by the research team. The experiment was conducted for four days in Tokyo and four days in Hiroshima. Schedules were on average 8-hours long and were designed considering three factors as shown in Table 1, namely, activity duration, designated floor area ratio (FAR) at destination (as defined in the city master plans,) and group size. FAR is used as a measure of building density, which is expected to affect accurate detection.

Such a design allows us to measure the potential of GLH data to identify joint activities under different conditions that might affect detection rates.

Each participant was provided with two Android phones and two iPhones and a GPS logger (see Table 2). Wi-Fi settings were conditionally randomized so that each participant had one Android phone and one iPhone with Wi-Fi on, and one Android phone and one iPhone with Wi-Fi off. Participants were required to carry all the equipment in a shoulder back provided by the research team.

A dummy Google account was created by the research team for each device used in the experiment. After signing into each device's Google account, Location History was turned on via the Google Account menu, only for the mobile device associated with the account.

Regarding device settings, for Android devices, in the Settings app's Security and Location menu, under the Mode submenu, the "High accuracy" setting was turned on. This setting uses GPS, Wi-Fi, mobile networks, and sensors to get the most accurate location as well as Google Location Services to estimate the device's location faster and more accurately (Google, 2022). For iPhones, in the Settings app's Location Services menu, Location Services was turned on. In the Google Maps' Location Information menu, permission was set to "Always" and the "Accurate information location" option was turned on.

In addition, participants were required to log online departure and arrival times to and from each location on real time via a Google form. This log constitutes the ground truth data of this study. The research team provided real-time support to participants via a chat app in order to respond efficiently to participants' inquiries, schedule changes, and to remind participants to log departure and arrival times in cases they have forgotten to do so. Since all Google accounts used in this experiment were dummy accounts created by the researchers, they had real-time access to the Location History of all participants and used this to verify, to the extent possible, that schedules were in fact being executed as planned. Participants were also informed that they would be monitored, as an incentive to execute the schedules carefully. Informal checks of location independent of GLH were the receipts for purchases conducted during the schedule execution, such as restaurant & café receipts, museum & garden ticket receipts etc.

Although GLH data can be edited by the user, in order to test GLH effectiveness as a completely passive data collection method, no editing was done to the data.

All experiments were conducted in December 2020. A pre-test survey was conducted with students recruited from Shibaura Institute of Technology in Tokyo.

**Table 1. Factors controlled for during the schedule design**

| Variable | Level | Definition | Additional explanation | Allocation Rule |
|---|---|---|---|---|
| Duration | 1 | 15 to 29 minutes | | Random allocation |
| | 2 | 30 to 44 minutes | | |
| | 3 | Over 45 minutes | | |
| Floor area ratio | 1 | Open space | | Assigned by rule |
| | 2 | Indoors - low density | FAR <300% | |
| | 3 | Indoors - mid density | 300% < FAR <700% | |
| | 4 | Indoors - high density | FAR >700% | |
| Group size | 1 | 1 person | | Assigned by rule |
| | 2 | 2 persons | | |
| | 3 | 3 persons | | |
| | 4 | 4 persons | | |
| Device | 1 | Android | | Available to all |
| | 2 | iPhone | | |
| Wi-fi setting | 1 | On | | Available to all |
| | 2 | Off | | |

**Table 2. Specifications of equipment provided to each participant**

| Equipment | Model/OS (confirmation pending) |
|---|---|
| Android phone 1 | Sharp Aquos sense basic 702SH, Android™ 8.0, Google maps latest version as of Dec. 5, 2020 |
| Android phone 2 | Kyocera Digno® -J, Android™ 8.1, Google maps latest version as of Dec. 5, 2020 |
| iPhone 1 | Apple iPhone XR, iOS13.1.2~14.0.1, Google maps latest version as of Dec. 5, 2020 |
| iPhone 2 | Apple iPhone 6s, iOS 12.4.1~14.1, Google maps latest version as of Dec. 5, 2020 |
| GPS logger | GNS 3000 |

The total number of observations collected is summarized in Table 3. Group-event count indicates the total number of activities executed by group size. Person-event count is the group-event count multiplied by the group size, while device-event count is the person-event count multiplied by four, or the number of devices each respondent carried during the experiment. The device-event count total is the number of observations used in the individual detection models described (all-devices case) in Section 6.2.1., while for the joint detection models, resampling was used to expand the sample size, as it will be explained in Section 6.2.2.

**Table 3. Number of observations collected during the experiment**

| Group size | Group-event count | Person-event count | Device-event count |
|---|---|---|---|
| 1 | 16 | 16 | 64 |
| 2 | 32 | 64 | 256 |
| 3 | 16 | 48 | 192 |
| 4 | 25 | 100 | 400 |
| **Total** | **89** | **228** | **912** |

## 3.2. Activity scheduling process

Based on the design characteristics defined in the previous section, we defined the activity schedules based on the following process:

1. Define total number of activities by group size

2. Define number of time blocks, experiment start/end locations, and start/end times
3. Randomly allocate activity duration times (as defined in Table 1)
4. Set FAR levels by time blocks to reduce travel time (Locations within the same FAR levels should be reachable on foot, thus imitating trip chaining of activities)
5. For each FAR level specify a district/area that meets the FAR criteria
6. Assign activities by group size (as defined in Step 1) to each time slot
7. Calculate actual departure and arrival times
8. Decide specific activity locations based on feasible travel times and adjust if necessary
9. If schedule cannot be solved within the specified experiment time window, remove one activity, and retry

### 3.3. COVID-19 infection prevention strategies

During the briefing sessions conducted every morning at the start of the experiment, participants' body temperature was checked. Participants were required to wear face masks and were provided with hand spray to disinfect frequently their hands. At the end of the experiment, all equipment was disinfected with alcohol.

Activity locations were set such that, following with government requests for COVID-19 infection prevention the three Cs (Closed spaces with poor ventilation, Crowded places with many people nearby, Close-contact settings such as close-range conversations). However, participants had the discretion to modify the schedule if based on their own judgement, a particular location did not meet infection prevention criteria.

The experiment was approved by the Research Ethics Committee of the Graduate School of Engineering, the University of Tokyo (Certificate KE20-61).

## 4. Google Location History & its data structure

Google Location History is a Google-account level setting that track the location of mobile devices that (i) are signed into a Google account, (ii) have Location History turned on and (iii) device settings allow Location Reporting (Google, 2022). Users have complete control on whether to turn on or off the Location History, with the default setting being off. They can also edit the Location History to correct detection errors. Users can also easily download their GLH data in JSON format, which makes it easy to process and analyse. As such, buying GLH data directly from users, instead of asking users to answer burdensome surveys, can be a feasible data collection alternative, provided accuracy levels meet the researcher's requirements.

As of December 2020, GLH data was structured as shown in Appendix 1. Generally speaking, GLH data is composed of "placeVisit" data, which refers to activities and "activitySegment" data which refers to travel. Since the focus of this study is joint activities, we focus on "placeVisit" data. For each "placeVisit" object, GLH provides information on location (such geographic coordinates, Google Place ID, location name and address,) activity duration (timestamps for starting and ending times,) as well as a measure of location confidence. We use two types of location information, namely, geographic coordinate location and Google place ID, which refers to a unique identifier for a location in the Google Places database and Google Maps (Google, 2021a). This is, to the best of our knowledge, the first study to evaluate the accuracy of Google Place ID.

Although the same information is provided for other candidate locations, in this study we used the data for the activity with the highest confidence level.

Activities are matched against the ground truth data by selecting the GLH activity which has the highest temporal overlap. Measures of spatial and temporal accuracy are then calculated for each ground-truth-GLH data pair, given predefined accuracy thresholds, as explained in the next subsections. Figure 1 plots an illustrative example of the ground truth data against Google Maps location history data.

Note that although location tracking rules are OS-dependent (Bays & Karabiyik, 2019), we can control for this difference without knowing the specifics of these rules by estimating accuracy levels of Androids and iPhones independently, and by adding an OS dummy variable to capture this effect in statistical models ( see Section 6.2.)

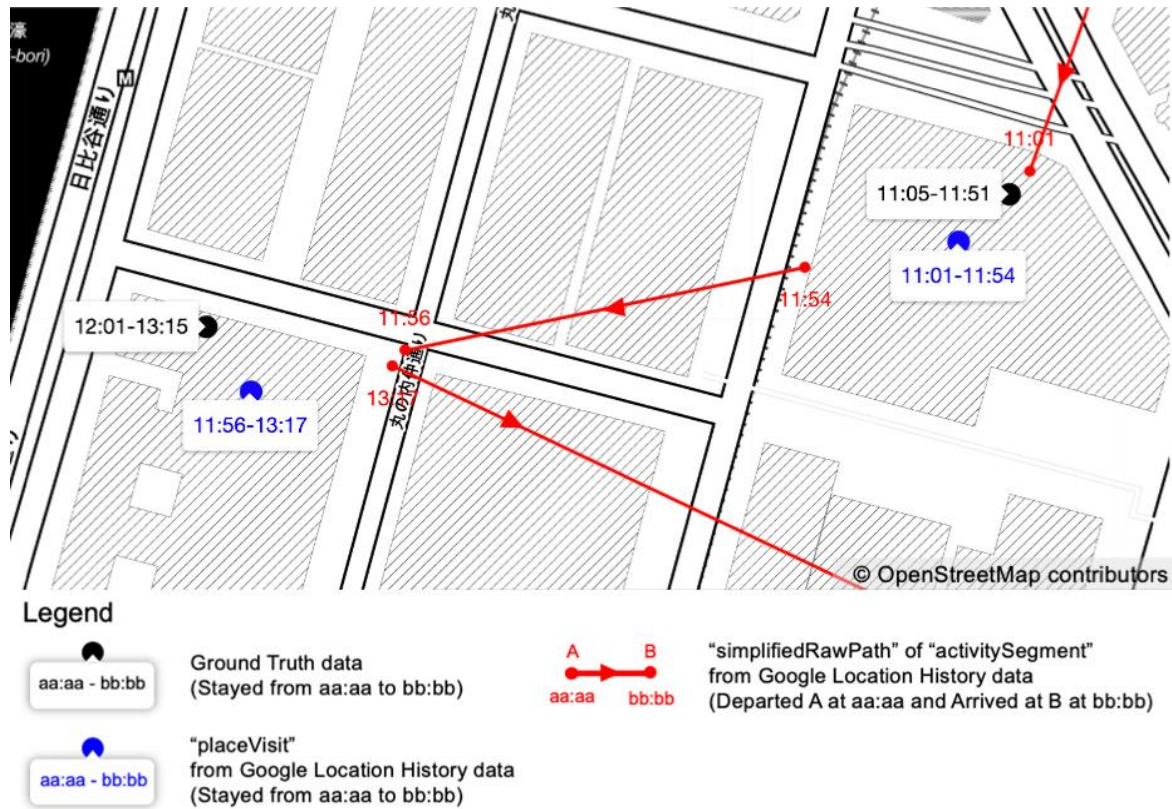

**Fig. 1. Ground truth data plotted against Google Maps Location History data.** GLH data records are composed of a "placeVisit" object reflecting places visited and an "activitySegment" object reflecting travel.

## 5. Measuring activity detection accuracy

Several accuracy measures have been proposed in the literature. Macarulla Rodriguez et al. (2018) used the accuracy radius (m) where it is possible to find the device at a given time, as given by GLH. If the GPS position (ground truth) was inside this radius, the observation was labeled a hit, and a miss otherwise.

Cools et al. (2021) matched the GLH data against the predefined travel diary (ground truth) by arrival and departure within an error of 10 minutes. Location position error was calculated as the Euclidean distance between GLH and ground truth. Arrival and departure times were evaluated using the root mean squared error (RMSE). Stay (dwell) time was evaluated using the normalized RMSE.

In this study we defined spatial and temporal accuracy measures independently to be able to evaluate detection accuracy at different spatial and temporal thresholds. This distinction is important given that it is difficult to identify optimal spatiotemporal thresholds, as they are dependent on the amount of error the analyst is willing to accept.

## 5.1. Spatial accuracy

Spatial accuracy *s* was measured as the Euclidean distance between the true location and the estimated location. The range of *s* is [0,∞) where 0 indicates perfect accuracy. For indoor locations, the coordinates of the centroid of the facility were used as a measure of the true location. For open spaces, which are extensive in area, and using the centroid would result in a larger error, a polygon of the perimeter was drawn, and distance was measured from the estimated location coordinates to the nearest point of the perimeter polygon. If the measurement was inside the polygon, accuracy is set to 0.

We also used Google Place ID match as a measure of accuracy, that is, whether the Google Place ID of the location defined in the activity schedule matches the Google Place ID identified by GLH. While the distance measure of accuracy described above only accounts for two dimensions as it cannot distinguish on what floor of a building the subject is, matching against Google Place Id overcomes this limitation by matching to a specific place or establishment. This also allows the analyst to retrieve location information stored in the Google Places database, which can potentially be used to infer activity purposes. However, in the case of a mismatch, this measure cannot identify how wrong the inference was in terms of how far this location was from the ground truth location.

## 5.2. Temporal accuracy

We defined three types of temporal accuracy measurements *t* (see Figure 2).

### 5.2.1. Intersect
Intersect refers to the temporal overlap between the ground truth schedule and GLH data. Activity start- and end-times (i.e., arrival and departure times) in the executed schedule were taken from the online log kept by participants and overseen by the researchers. Activity start- and end-times in the observed schedule were taken from the timestamps logged in the GLH data. The range of this measure is [0,1] where 1 indicates complete overlap between GLH and ground truth data. This measure, however, is not sensitive to errors in arrival and departure times in the cases the ground truth time period is completely covered by the GLH time period.

### 5.2.2. Divergence error
Divergence error refers to a divergence in start-or end-times between the GLH data and the ground truth. The range of this measure is [0,∞).

### 5.2.3. Missing observation
Refers to the case where no activity was recorded by GLH in spite of the occurrence of one. It is defined as a binary variable that takes value 1 if the intersect equals zero, and 0 otherwise.

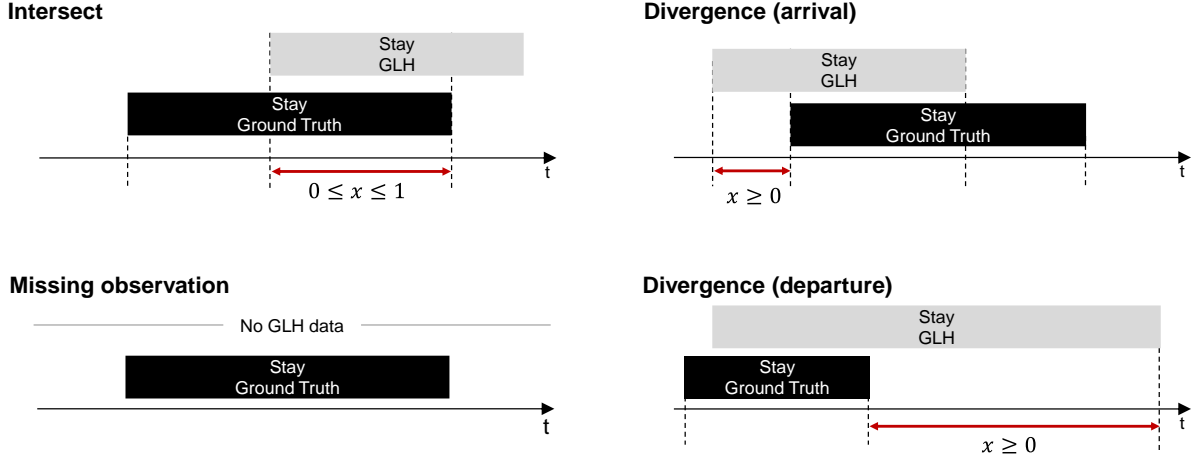

**Fig. 2. Illustration of four different measures of temporal accuracy**

For the present analysis we mainly focus on the intersect measure as a temporal accuracy measure, while accounting for the number of missing observations.

## 5.3. Activity detection rate

To define the activity detection rate, we first define an activity $a = (x_a, J_a)$ where $x_a$ is a vector of ground truth data activity attributes (including Google Place ID $i_a$), and $J_a$ is the set of observations of individuals who participated in activity $a$. An individual observation $j \in J_a$ is defined as $j = (w_j, d_j, i_j, s_{ja}, t_{ja})$ : $w_j$ and $d_j$ are respectively wi-fi setting and device of $j$, $i_j$ is the identified Google Place ID for $j$, and $s_{ja}$ and $t_{ja}$ are measures of spatial and temporal accuracy. An individual activity detection $\delta_j(S, T)$ of $j$ within a spatial accuracy threshold $S$ and a temporal accuracy threshold $T$ is defined as

$$\delta_j(S,T) = \begin{cases} 1 & if\ s_{ja} \geq S\ and\ t_{ja} \leq T \\ 0 & otherwise \end{cases} \quad (1)$$

Note that, when Google Place ID match is used as a spatial accuracy measure instead of Euclidean distance, it is replaced by

$$\delta_j^{gID}(T) = \begin{cases} 1 & if\ i_j = i_a\ and\ t_{ja} \leq T \\ 0 & otherwise \end{cases} \quad (2)$$

A group activity detection $\Delta_a(S, T)$ is then defined as the product of individual detections:

$$\Delta_a(S,T) = \prod_{j \in J_a} \delta_j(S,T) \quad (3)$$

Finally, we obtain the activity detection rate for $g$-group-sized activities $P_g(S, T)$ within spatial accuracy threshold $S$ and temporal accuracy threshold $T$, which is defined as

$$P_g(S,T) = N_g^{-1} \sum_{a, |J_a|=g} \Delta_a(S,T) \quad (4)$$

where $N_g$ indicates the total number of $g$-group-sized activity observations[1].

# 6. Results

## 6.1. Aggregate results

Figure 3 summarizes the detection rates given different spatiotemporal threshold levels, devices, and group sizes. The x-axis shows different thresholds of temporal accuracy T, which is the minimum acceptable overlapping between ground truth and GLH data. The y axis shows different thresholds of spatial accuracy S, which is the maximum acceptable Euclidean distance between the ground truth location and the GLH location. The closer to the bottom left corner of the figure, the stricter the spatiotemporal thresholds are. The bottom row shows the accuracy results using the Google Place ID match. Different colored bars indicate the group size.

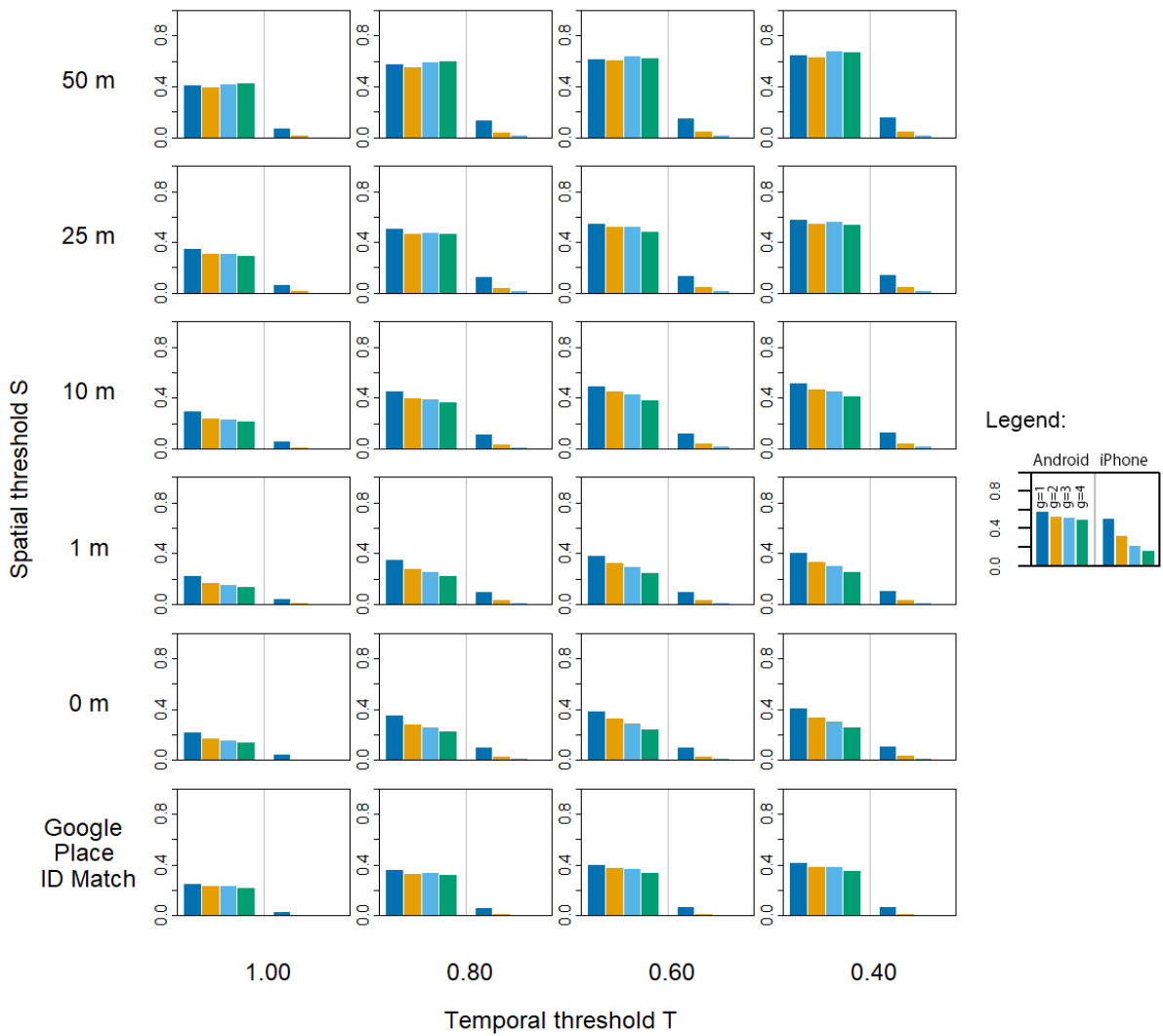

**Figure 3: Aggregate spatiotemporal accuracy given different accuracy thresholds, devices, and group sizes. The x-axis shows different thresholds of temporal accuracy T, which is the minimum acceptable**

---

[1] This is the number of observations recorded across all devices, not the number of ground truth activities. For example, since all participants carried four mobile devices, one 4-group-size activity is equivalent to 16 activity observations.

**overlapping between ground truth and GLH data. The y axis shows different thresholds of spatial accuracy S, which is the maximum acceptable Euclidean distance between the ground truth location and the GLH location. For each plot, the left panes show Android accuracy levels while the right panes show iPhone accuracy levels. Colored bars indicate different group sizes *g*.**

The first issue to identify is the stark difference in detection rates between iPhones and Android devices, irrespective of thresholds. While direct comparisons are difficult due to methodological differences, differences in accuracy levels have also been reported by Cools et al. (2021).

For Android devices, using Google Place ID match at $T = 1.00$ (the strictest threshold setting,) activity detection rates ranged from 22% to 25.7% for $g = 4$ and $g = 1$, respectively. As expected, relaxing the accuracy thresholds yields higher detection rates. For example, In the case of $S = 10m$ and $T = 0.8$, detection rates ranged from 37.2% to 45.6%, for $g = 4$ and $g = 1$, respectively. Further relaxing the spatial threshold to a less strict, yet still operational threshold of 50m, detection rates approached 60%. It must be noted however, that it is hard to identify what the optimal values for thresholds $S, T$ are since it would usually depend on the objectives of the study and the amount of error the analyst is willing to accept.

As hypothesized, larger groups result in lower detection rates, but these reductions are smaller than we expected.

## 6.2. Modelling factors affecting detection probability

To evaluate the extent to which several factors affect detection probability, we estimated two sets of binary logit models given different spatiotemporal accuracy thresholds. The first set of models focuses on detection probability of individual devices, while the second set of models focuses on joint detection probability. Table 4 summarizes the definition of the explanatory variables used in the individual detection and joint detection models.

Table 4: Definition of explanatory variables

| Variable | Definition | Unit | Used in individual detection model | Used in joint detection model |
| --- | --- | --- | --- | --- |
| Floor area ratio (FAR) | FAR at activity location as designated on city master plan FAR refers to the ratio between a building's total area and the lot where the building is located | FAR/100 | Yes | Yes |
| Activity duration | Total duration of activity as measured in the ground truth | Minute | Yes | Yes |
| Open space | Dummy variable. Takes value 1 if activity location is an open space, 0 otherwise | Dummy | Yes | Yes |
| Android device | Dummy variable. Takes value 1 if device is Android, 0 otherwise | Dummy | Yes (all devices model only) | No |
| Android ratio | Ratio of devices in group that are Androids | Ratio | No | Yes (all devices model only) |
| Android 702SH model | Dummy variable. Takes value 1 if device is Sharp Aquos 702SH , 0 otherwise | Dummy | Yes (Android model only) | No |
| Android 702SH model ratio | Ratio of devices in group that are Sharp Aquos 702SH | Ratio | No | Yes (Android model only) |

| Wi-fi on | Dummy variable. Takes value 1 if Wi-fi is turned on, 0 otherwise | Dummy | Yes | No |
|---|---|---|---|---|
| Wi-fi on ratio | Ratio of devices in group that have Wi-fi turned on | Ratio | No | Yes |
| Group size | Number of activity participants | Number of persons | No | Yes |

### 6.2.1. Factors affecting detection of individual devices

The first set of models focuses on detection probability of individual devices (n=912). The dependent variable takes value 1 if the ground truth activity was detected given thresholds $S, T$, and 0 otherwise. For the individual devices case, we estimated effect sizes and confidence intervals using the bootstrapping method with 300 iterations per model. At each iteration elasticities and marginal effects for continuous variables were estimated analytically and aggregated using the probability-weighted sample enumeration method, while marginal effects for dummy variables were estimated via simulation. Figure 4 summarizes elasticities and marginal effects of variables affecting detection probability at the individual level given different spatiotemporal accuracy thresholds.

Floor area ratio[2] (FAR) at destination is negatively associated with detection probability. Estimated elasticities suggest a non-trivial effect size, irrespective of the spatiotemporal threshold used. For example, for $S = 10m$ and $T = 0.8$ the elasticity point estimate is -0.45, suggesting reduction of 0.45% in detection probability given a 1% increase in FAR/100. The effect is substantially larger when matching against Google Place ID, with point estimates of -0.64 and -0.63 for $T = 0.6$ and $T = 0.8$, respectively.

As hypothesized, activity duration is positively associated with higher detection probability. For $S = 10m$ and $T = 0.8$, the elasticity point estimate is 0.75, indicating a 0.75% increase in detection probability given a 1% increase in activity duration. However, when matching against Google Place ID, not only the point estimates are considerable smaller, but these are estimated with higher uncertainty, taking negative values near the confidence interval lower bounds.

Consistent with the detection accuracy results presented in the previous section, using an Android device is positively associated with detection probability. For $S = 10m$ and $T = 0.8$, the estimated marginal effect indicates an average increase of 34 percentage points in detection probability over an iPhone, a substantial effect. A similar magnitude is observed when matching against Google Place ID, with average increases in detection probability of 33 and 30 percentage points for $T = 0.6$ and $T = 0.8$ cases, respectively.

In terms of Wi-Fi settings (on), the marginal effect point estimates are small and exhibit a high degree of uncertainty irrespective of the spatiotemporal threshold used, with all confidence intervals including zero. When considering both point estimates and estimate uncertainty, a clear and non-trivial effect on individual detection probability cannot be identified.

Finally, regarding the effects of open space (whether the destination is an open space or not), findings are mixed. While for $S = 10m$ and $T = 50m$, the effects are relatively small, and in some cases the estimates have high uncertainty (the confidence intervals of the $T =$

---

[2] During the estimation process, the effect of mobile antenna density was also tested. This variable is however highly correlated with FAR so its effect cannot be properly captured. Furthermore, we tested a model antenna density and excluding FAR variable, but it underperformed. Hence, this variable was not included in the final analysis.

0.6 cases include zero), open space is negatively associated with detection probability. Furthermore, when matching against Google Place ID, marginal effects point estimates suggest average decreases of 20 and 22 percentage points in detection probability, for $T = 0.6$ and $T = 0.8$, respectively. These effects are opposite to our hypothesis. We further tested different effect interactions and segmentation by area of open space, to test whether we could identify the reason for the observed effect direction, and in all cases estimated effects were consistently negative. A potential explanation for this phenomenon is that modern devices depend less on GPS data (which one would expect to perform better in open spaces) and use additional information from other sources such as Wi-Fi, mobile networks, and device sensors (Google, 2021), that might favor detection near or inside buildings to some extent. However, further analysis is necessary to validate this hypothesis.

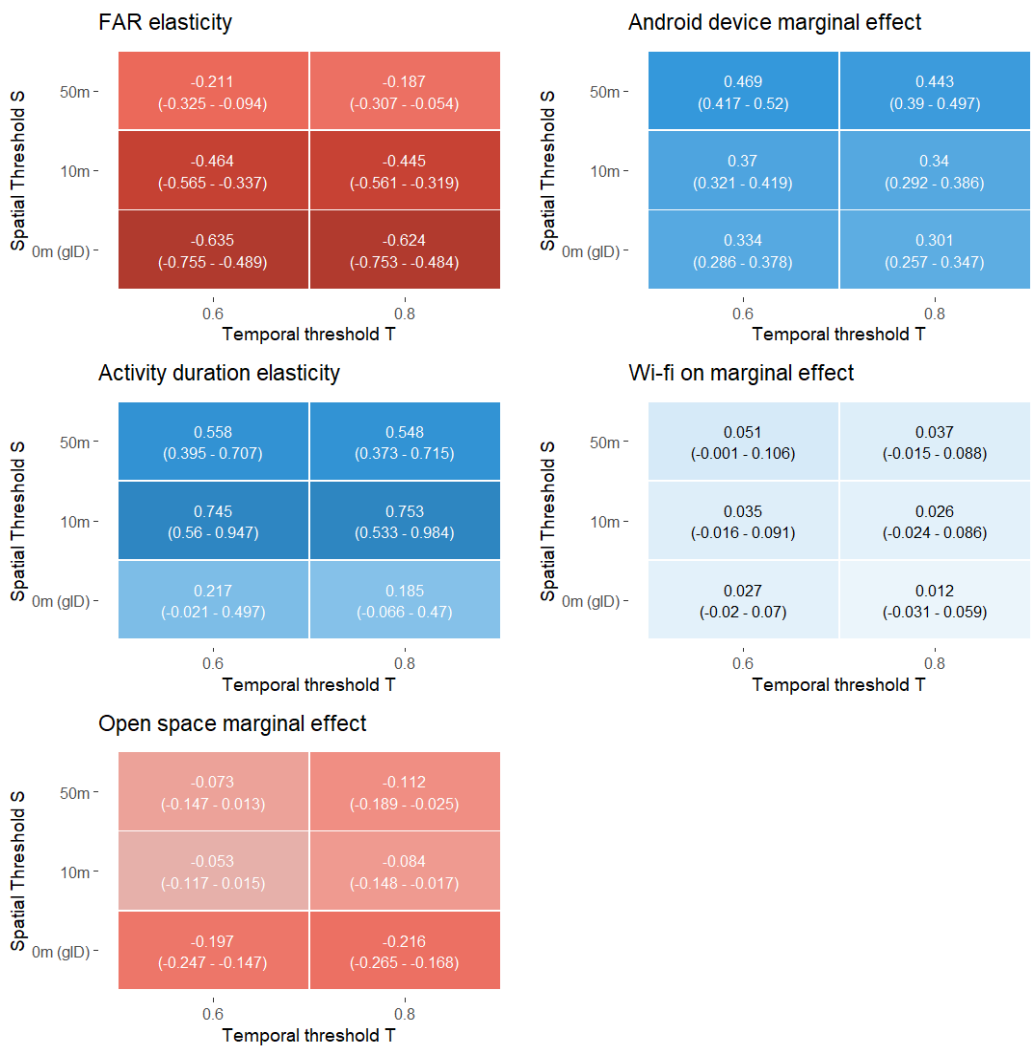

**Figure 4: Effect magnitude of variables affecting detection at the individual level given different spatiotemporal accuracy thresholds. Values in parenthesis show the respective 95% confidence intervals.**

### 6.2.2. Factors affecting joint detection probability
The second set of models evaluate which factors affect detection at the group level. To do so, we generated composite 2-person, 3-person, and 4-person groups by resampling from all possible permutations from the observed two-person, three-person and four-person groups. For example, from all the observed 4-person group activities (for each activity, 4 persons x 4 devices = 16 devices) we generated all possible 2-person, 3-person, and 4-person permutations. From these permutations, we randomly sampled an equal number of samples for all possible permutations of "wi-fi on ratio", "Android ratio" and "group size". This was done to mitigate potential biases that might remain after the scheduling process described in Section 3.2. Sample size for these models was 3296. The sampling process was repeated 300 times and estimated effect magnitudes were averaged over all iterations (at each iteration, elasticities and marginal effects were calculated in the same ways in the individual devices case). 95% confidence intervals were estimated empirically by calculating the 2.5 and 97.5 percentiles of the empirical distribution of estimated effects.

Figure 5 summarizes elasticities and marginal effects of variables affecting joint activity detection probability given different spatiotemporal accuracy thresholds. Similar to the individual detection case, the elasticities of FAR suggest a non-trivial effect magnitude on joint activity detection probability, with effect magnitude growing given stricter spatial thresholds. For $T = 0.8$, point estimates are -0.18, -0.45 and -0.73 for $S = 50m$, $S = 10m$ and Google Place ID match, respectively.

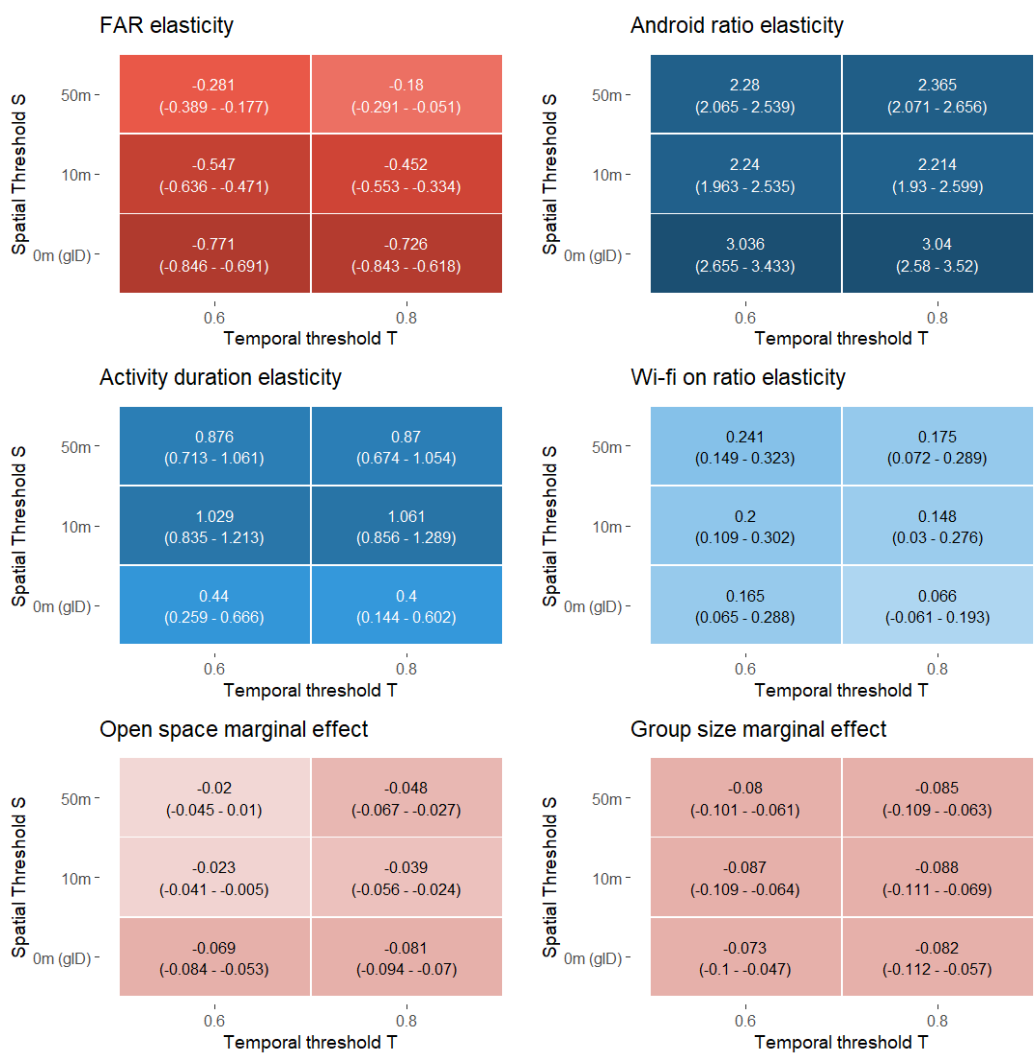

**Figure 5: Effect magnitude of variables affecting joint detection given different spatiotemporal accuracy thresholds. Values in parenthesis show the respective 95% confidence intervals.**

Activity duration is positively associated with increases in joint detection probability, irrespective of spatiotemporal threshold. For $S = 10m$ and $T = 0.8$, the elasticity point estimate is 1.06, indicating a 1.06% increase in detection probability given a 1% increase in activity duration, a substantial effect. When matching against Google Place ID, while smaller than the $S = 50m$ and $S = 10m$ cases, the effect of activity duration is still relatively large, with point estimates of 0.44 and 0.40 for $T = 0.6$ and $T = 0.8$, respectively. The difference in effect magnitudes, however, does suggest that matching against Google Place ID is less sensitive to activity duration.

The ratio of Android devices in the group is positively associated with detection and had the largest effect magnitude of all variables, irrespective of spatiotemporal threshold. For $S = 10m$ and $T = 0.8$, the elasticity point estimate indicates an average 2.21% increase in detection probability given a 1% increase in Android device ratio. When matching against Google Place ID, elasticity point estimates are 3.04 for both $T = 0.6$ and $T = 0.8$. These effects are consistent with the accuracy estimates reported in Fig. 3.

In spite of the weak effects measured for individual device detection, the ratio of devices with Wi-Fi turned on is in general positively associated with joint detection probability. For $S = 10m$ and $T = 0.8$, the elasticity point estimate indicates an average 0.15% increase in joint detection probability, a moderate effect. Note however, that when matching against Google Place ID, for $T = 0.8$, the confidence interval includes zero.

As hypothesized, group size is negatively associated with detection probability, and its effect magnitude is very consistent across different spatiotemporal thresholds. For $S = 10m$ and $T = 0.8$, an additional member in the party is associated on average with an 8.8 percentage point reduction in joint detection probability. This means a 26.4 percentage point difference in detection probability between an individual activity and a 4-person joint activity, a non-trivial effect size.

Finally, and similarly to the individual detection case, open space is negatively associated with detection probability, although the effect magnitude is considerably smaller. This reduction is particularly large when matching against Google Place ID. For $S = 10m$ and $T = 0.8$, the marginal effect point estimate suggests that conducting an activity in an open space reduces joint detection probability by 4 percentage points on average. When matching against Google Place ID, estimated average reductions of joint detection probability are of 7 and 8 percentage points for $T = 0.6$ and $T = 0.8$, respectively.

### 6.2.3. Android-only estimates

Given the stark differences in accuracy between Android and iPhone devices discussed earlier, we conducted the same analysis as in the previous section for the Android-only case (n=456). Figures 6 and 7 summarize effect magnitudes for individual detection and joint detection, respectively. We will focus the discussion mainly on the key differences identified. We also tested whether or not the type of Android device affected detection probability.

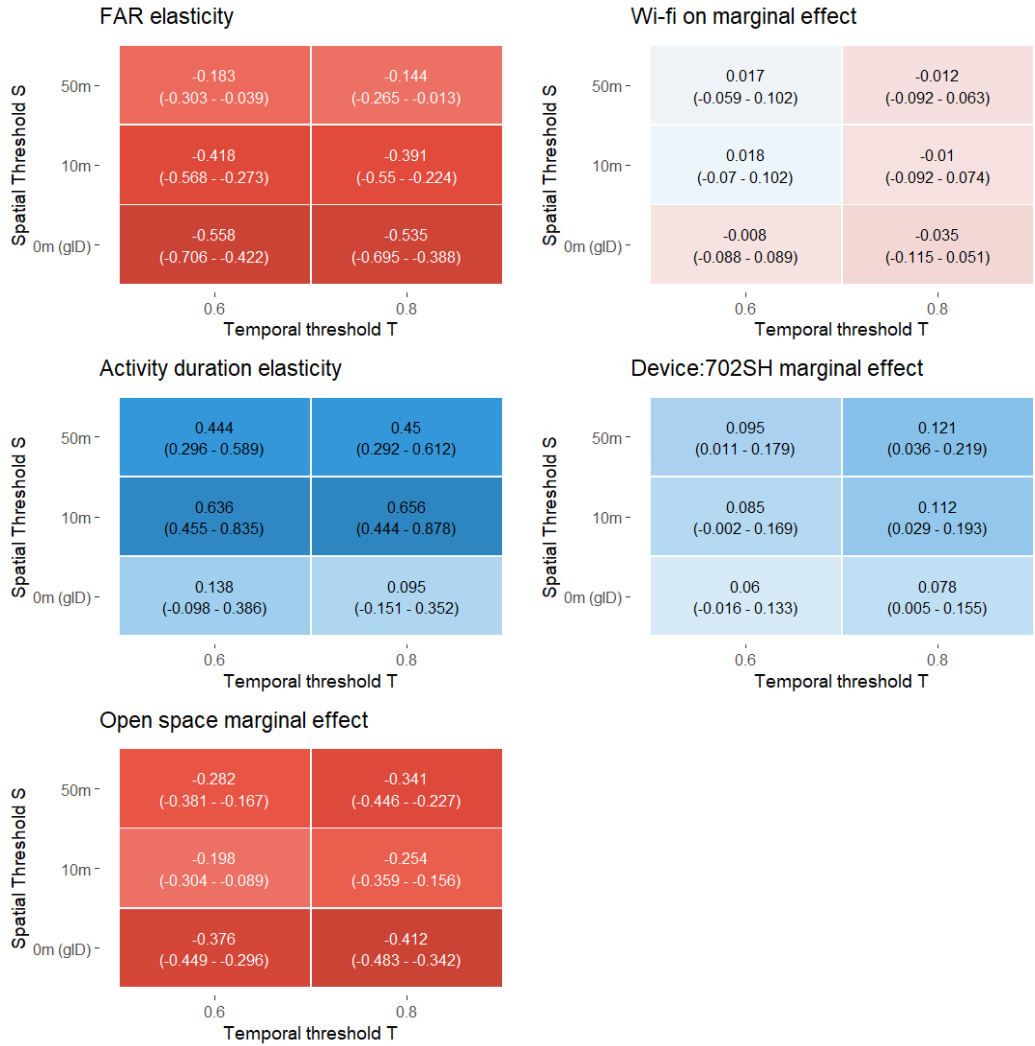

**Figure 6: Effect magnitude of variables affecting detection at the individual level given different spatiotemporal accuracy thresholds. Values in parenthesis show the respective 95% confidence intervals. Android devices only.**

For the individual detection case, effect directions are consistent with the all-devices case, and similar trends can be observed in terms of effect magnitude. The key difference is that the magnitude of the open space effect on detection probability is considerably higher for the Android-only case. We also found evidence that device type affected detection probability. In this study we used two different Android devices (Sharp Aquos Sense 702SH and Kyocera Digno-J). For the individual detection case we added a dummy variable for Sharp Aquos 702SH devices (hereinafter 702SH), to capture its effect difference on detection probability. All else equal, compared to the Kyocera devices, 702SH devices are positively associated with higher detection probability. For $S = 10m$ and $T = 0.8$, the marginal effect point estimate suggests an increase in detection probability, of 11.2 percentage points, and 7.8 percentage points when matching against Google Place ID. At the same time, some estimates have higher uncertainty, with two instances (for $T = 0.6$, $S = 10m$ and Google Place ID match,) where zero is included in the confidence intervals.

Regarding joint detection (n=474), effect directions are also consistent with the all-devices case. FAR and activity duration effect magnitudes are in general smaller, but the overall trend is similar. However, for the remaining variables, some important differences are worth mentioning. First, similar to the individual detection case, the magnitude of the open space effect is larger for the Android-only case irrespective of spatiotemporal accuracy threshold.

Contrary to what was observed for the all-devices case, the ratio of devices with Wi-Fi is negatively associated with joint activity detection probability. For $T = 0.8$, irrespective of spatial thresholds, the effect was negative with point elasticities of -0.10, -0.12 and -0.19 for $S = 50m, 10m$ and Google Place ID match, respectively. That being said, estimated effects are smaller and have higher uncertainty for the $T = 0.6$ cases, where all confidence intervals include zero. Regardless, the contrasting results between the all-devices model and the Android only models suggests that Wi-Fi effects might differ between iPhones and Androids. Given that Wi-Fi is one of the sources used to get the most accurate location (Google, 2022), more research is needed to properly clarify the effects of using Wi-Fi on location accuracy on different devices.

Also differing from what was observed for the all-devices case, no clear effect can be identified for group size on joint detection accuracy, with marginal effect point estimates close to zero for all cases. One possibility is that group size will not substantially affect joint activity detection, and that detection probability is location specific in an all-Android setting. If that is the case, we might be able to capture group activities even for larger group size for particular locations where these detection probabilities are relatively high.

Finally, regarding the elasticity of the ratio of 702SH devices, a positive association with joint detection is observed, irrespective of spatio-temporal threshold. For $S = 10m$ and $T = 0.8$, the elasticity point estimate is 0.23, indicating a 0.23% increase in detection probability given a 1% increase in the ratio of 702SH devices, a non-negligible effect. When matching against Google Place ID, a similar magnitude is observed, with an elasticity point estimate of 0.21. This is consistent with the individual detection case, as evidence that device type affects detection probability. Although the number of devices we tested is limited, these findings suggest that attention should be paid to device accuracy when conducting a study.

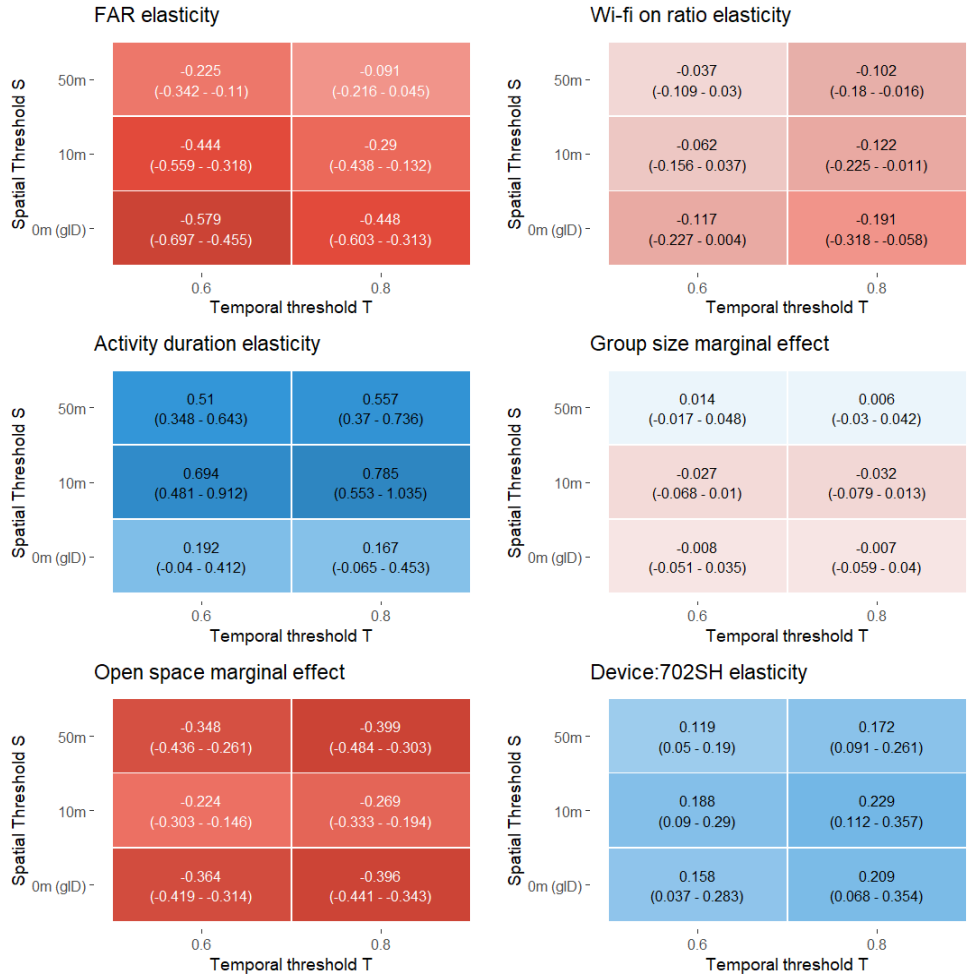

**Figure 7:** Effect magnitude of variables affecting joint detection given different spatiotemporal accuracy thresholds. Values in parenthesis show the respective 95% confidence intervals. Android devices only.

#### 6.2.4. Predictive accuracy of estimated models

To evaluate the prediction accuracy of estimated models, internal validation was conducted via 10-fold cross-validation. This was done to avoid the "optimism" of in-sample goodness of fit statistics (Parady, Ory and Walker, 2021). Tables 5-8 summarize the cross-validation results for (i) all devices and Android-only cases, (ii) individual and joint detection, and for (iii) different spatiotemporal thresholds.

The first issue to point out is the clear difference between true positive rates (TPR) and true negative rates (TNR) across all models. While both TNR and TPR (as well as other measures derived from these statistics) are important accuracy measures, the high observed TNR for these models are related to the observed (ground truth) detection rates. That is, there are a lot of negative outcomes (especially in the joint-detection case), which makes them easier to predict. As such, focusing exclusively on model accuracy or TNR might be misleading. This is also true for rho-square measures, which are usually dependent on base detection rates. As such, we will base our discussion on the true positive rates, which in this case provide a better picture of model accuracy.

For type 1 models ($S = gID$, $T = 0.8$), which are the models that use the strictest spatiotemporal thresholds in Tables 5-8, true positive rates are 36.8% (individual detection, all devices), 47.6% (individual detection, Android-only), 9.8% (joint detection, all devices), and 29.1% (joint detection, Android-only). As expected, true positive rates are considerably lower for joint activities. While we have identified several factors associated with detection and quantified its effect magnitudes, there is certainly room for improvement in terms of identifying other factors associated with detection, both at the individual and at the joint level.

**Table 5. 10-fold cross-validation results for activity detection models at the individual level (all devices)**

| Measure | Model 1 | Model 2 | Model 3 | Model 4 | Model 5 | Model 6 |
|---|---|---|---|---|---|---|
| Threshold settings | $S = gID$ $T = 0.8$ | $S = gID$ $T = 0.6$ | $S = 10m$ $T = 0.8$ | $S = 10m$ $T = 0.6$ | $S = 50m$ $T = 0.8$ | $S = 50m$ $T = 0.6$ |
| Num. observations | 912 | 912 | 912 | 912 | 912 | 912 |
| GLH activity detection rate(observed) | 21.8% | 23.7% | 28.6% | 31.0% | 36.2% | 38.8% |
| Model accuracy | 79.7% | 78.5% | 76.3% | 76.4% | 75.6% | 74.8% |
| Model balanced accuracy (TPR+TNR)/2 | 64.2% | 64.4% | 65.8% | 68.7% | 73.1% | 73.7% |
| TPR (True positive rate) | 36.8% | 37.8% | 41.5% | 48.6% | 64.4% | 69.0% |
| TNR (True negative rate) | 91.6% | 91.0% | 90.1% | 88.7% | 81.8% | 78.3% |
| PPV (Positive prediction value) | 10.2% | 11.6% | 15.7% | 19.9% | 30.9% | 35.9% |
| FNR (False negative rate) | 63.2% | 62.2% | 58.5% | 51.4% | 35.6% | 31.0% |
| FPR (False positive rate) | 8.4% | 9.0% | 9.9% | 11.3% | 18.2% | 21.7% |
| Rho-square | 0.40 | 0.38 | 0.29 | 0.29 | 0.26 | 0.25 |
| Adjusted rho square | 0.39 | 0.37 | 0.28 | 0.27 | 0.24 | 0.24 |

**Table 6. 10-fold cross-validation results for activity detection models at the individual level (Android only)**

| Measure | Model 1 | Model 2 | Model 3 | Model 4 | Model 5 | Model 6 |
|---|---|---|---|---|---|---|
| Threshold settings | $S = gID$ $T = 0.8$ | $S = gID$ $T = 0.6$ | $S = 10m$ $T = 0.8$ | $S = 10m$ $T = 0.6$ | $S = 50m$ $T = 0.8$ | $S = 50m$ $T = 0.6$ |
| Num. observations | 456 | 456 | 456 | 456 | 456 | 456 |
| GLH activity detection rate (observed) | 36.8% | 40.3% | 45.6% | 49.6% | 58.3% | 62.3% |
| Model accuracy | 66.0% | 61.8% | 64.7% | 62.3% | 66.6% | 67.1% |
| Model balanced accuracy (TPR+TNR)/2 | 62.5% | 59.3% | 64.1% | 62.4% | 64.6% | 62.2% |
| TPR (True positive rate) | 47.6% | 45.6% | 55.7% | 58.8% | 77.6% | 82.6% |
| TNR (True negative rate) | 77.3% | 73.1% | 72.5% | 66.0% | 51.6% | 41.8% |
| PPV (Positive prediction value) | 26.2% | 29.6% | 39.2% | 46.7% | 68.0% | 76.5% |
| FNR (False negative rate) | 52.4% | 54.4% | 44.3% | 41.2% | 22.4% | 17.4% |
| FPR (False positive rate) | 22.7% | 26.9% | 27.5% | 34.0% | 48.4% | 58.2% |
| Rho-square | 0.17 | 0.12 | 0.10 | 0.09 | 0.15 | 0.14 |
| Adjusted rho square | 0.15 | 0.10 | 0.08 | 0.07 | 0.12 | 0.12 |

**Table 7. 10-fold cross-validation results for joint activity detection models (all devices)**

| Measure | Model 1 | Model 2 | Model 3 | Model 4 | Model 5 | Model 6 |
|---|---|---|---|---|---|---|
| **Threshold settings** | $S = gID$ $T = 0.8$ | $S = gID$ $T = 0.6$ | $S = 10m$ $T = 0.8$ | $S = 10m$ $T = 0.6$ | $S = 50m$ $T = 0.8$ | $S = 50m$ $T = 0.6$ |
| **Num. observations** | 3296 | 3296 | 3296 | 3296 | 3296 | 3296 |
| GLH activity detection rate (observed) | 6.3% | 8.9% | 9.3% | 11.7% | 11.7% | 14.5% |
| Model accuracy | 94.0% | 91.5% | 91.0% | 89.1% | 89.6% | 87.4% |
| Model balanced accuracy (TPR+TNR)/2 | 54.7% | 58.5% | 56.6% | 61.0% | 61.0% | 65.2% |
| TPR (True positive rate) | 9.8% | 18.3% | 14.3% | 24.3% | 23.6% | 33.8% |
| TNR (True negative rate) | 99.5% | 98.7% | 98.9% | 97.7% | 98.4% | 96.5% |
| PPV (Positive prediction value) | 0.7% | 1.7% | 1.4% | 3.2% | 3.0% | 5.6% |
| FNR (False negative rate) | 90.2% | 81.7% | 85.7% | 75.7% | 76.4% | 66.2% |
| FPR (False positive rate) | 0.5% | 1.3% | 1.1% | 2.3% | 1.6% | 3.5% |
| Rho-square | 0.76 | 0.70 | 0.67 | 0.63 | 0.62 | 0.58 |
| Adjusted rho square | 0.75 | 0.69 | 0.67 | 0.63 | 0.61 | 0.57 |

**Table 8. 10-fold cross-validation results for joint activity detection models (Android only)**

| Measure | Model 1 | Model 2 | Model 3 | Model 4 | Model 5 | Model 6 |
|---|---|---|---|---|---|---|
| **Threshold settings** | $S = gID$ $T = 0.8$ | $S = gID$ $T = 0.6$ | $S = 10m$ $T = 0.8$ | $S = 10m$ $T = 0.6$ | $S = 50m$ $T = 0.8$ | $S = 50m$ $T = 0.6$ |
| **Num. observations** | 474 | 474 | 474 | 474 | 474 | 474 |
| GLH activity detection rate (observed) | 29.5% | 41.1% | 37.8% | 47.3% | 50.2% | 60.3% |
| Model accuracy | 70.0% | 64.8% | 67.5% | 64.4% | 71.5% | 70.5% |
| Model balanced accuracy (TPR+TNR)/2 | 58.4% | 62.3% | 63.2% | 63.9% | 71.3% | 67.8% |
| TPR (True positive rate) | 29.1% | 49.8% | 45.5% | 54.9% | 78.4% | 79.8% |
| TNR (True negative rate) | 87.6% | 74.9% | 81.0% | 72.8% | 64.2% | 55.7% |
| PPV (Positive prediction value) | 12.2% | 31.6% | 25.5% | 40.4% | 55.0% | 68.4% |
| FNR (False negative rate) | 70.9% | 50.2% | 54.5% | 45.1% | 21.6% | 20.2% |
| FPR (False positive rate) | 12.4% | 25.1% | 19.0% | 27.2% | 35.8% | 44.3% |
| Rho-square | 0.27 | 0.14 | 0.18 | 0.12 | 0.21 | 0.18 |
| Adjusted rho square | 0.25 | 0.12 | 0.15 | 0.10 | 0.18 | 0.16 |

# 7. Executing an empirical study

Although not the main subject of this study, in this section we detail how an empirical study could be executed using GLH as well as discuss some potential limitations.

A survey that considers the travel behavior of group members presents some logistical difficulties that explain why such studies have not been, to the best of our knowledge, executed. Without a priori knowledge of social networks or group composition, in order to identify groups, a snowball sampling method is required, where an initial sample, or seeds, are randomly sampled, and in addition to basic sociodemographics, are asked to identify their social networks, usually via a name generator. Identified network members will in turn be recruited to the survey. Precedents exist in the transportation field, such as the study by Kowald and Axhausen (2012) that used snowball sampling method to collect information on Swiss social networks. However, such studies impose high response burden on participants. The key merit of using GLH is that one can directly purchase GLH data from sampled individuals or households to collect travel behavior data, thus drastically reducing the response burden.

GLH data collection can be gathered using two approaches, (i) buying GLH data as-is for a time period prior to the survey execution, or (ii) buying GLH data for a determined time period starting after recruitment. The first approach has the lowest response burden, since it only requires users to download their own GLH data, a rather straightforward process. As of December 2020, GLH data was stored for a maximum of 18 months. The downside of this approach is that researchers have no control of Google Maps or device settings that might affect accuracy, or whether or not GLH was activated for any given period of time. The second approach overcomes to some extent these limitations since researchers can instruct and assist users to have the desired settings. Furthermore, GLH data on activity location, time, and transportation mode can be edited by users. Google Maps Timeline displays the inferred activities, based on which users easily edit the GLH data. Of course, this editing imposes an additional burden; however, it is less burdensome than filling traditional activity-diary surveys. Editing might be required particularly for iPhone users, to overcome the limitations of low detection accuracy.

There are several limitations to using GLH data. First, it will require certain degree of IT literacy for participants to download their own GLH data and share it with researchers. This will bias the sample towards younger participants, although we expect this bias to be mitigated over time as younger generations age. Second, although monetary incentives can be used to increase participation rates, researchers must persuade users to participate not only by explicitly explaining the purposes of the study, but also how their privacy is protected and how the collected data will be handled during and after the research project is over. The self-selection issue, although pervasive in all types of surveys, might be exacerbated by privacy concerns, which might be different across cultural and socio-political contexts.

Finally, snowball sampling is not a probabilistic sampling method, so the generalizability of the findings is limited. On the other hand, irrespective of the representativeness of the data, collecting long-term data on groups' travel behavior might allow for methodological improvements that have been limited so far by data availability.

Once data is collected, Place ID can be used to extract location information from the Google Places database and infer activity purpose.

Since detection accuracy is context-dependent, an empirical application will ideally have a subset of the sample for which ground truth data is also collected (i.e., A travel diary, or digital alternatives via specialized apps,) and used to evaluated accurate detection probability, in a similar way to what we have shown in this study.

In addition, for joint activity simulation analysis, detection probability models can be estimated and used to get a joint activity detection propensity score. Based on this score, the inverse probability weighting method (Wooldridge, 2007) could be utilized to obtain an unbiased frequency of joint activities for a particular group from GLH data. Provided a large enough dataset is collected the characteristics of the missing activities can be potentially inferred from the distributions of the observed data, although ideally, these characteristics would be predicted based on a joint activity generation model. The details of such model are, however, out of the scope of this study. At any rate, given present levels in activity detection rates, GLH data will be most useful in mid- to long-term observation periods, given the high spatiotemporal variability of social activities.

## 8. Discussion and conclusion

This study evaluated the potential of using Google Maps Location History data to detect joint activities in social networks. For Android devices, using Google Place ID match at $T = 1.00$ (the strictest threshold setting,) activity detection rates ranged from 22% to 25.7% for $g = 4$ and $g = 1$, respectively. In the case of $S = 10m$ and $T = 0.8$, detection rates ranged from 37.2% to 45.6%, for $g = 4$ and $g = 1$, respectively. Further relaxing the spatial threshold to a less strict, yet still operational threshold of 50m, detection rates approached 60%. Less strict thresholds resulted in higher detection rates, but with higher spatiotemporal error.

Logit models were estimated to evaluate factors affecting activity detection. Generally, in terms of magnitude, non-trivial effects were found for floor area ratio (FAR) at location, activity duration, Android device ratio, device model ratio, whether the destination was an open space or not, and group size on joint activity detection probability.

While to some extent we agree with the conclusion reached by Cools et al. (2021) that current detection rates might limit its usefulness in travel behavior studies, we argue that these detection rates, if not ideal, must be weighed against the potential of observing travel behavior and joint activities over long periods of time, a longstanding limitation of the field. Furthermore, GLH data could potentially be used in conjunction with other data-gathering methodologies to compensate for some of its limitations. For example, we could utilize the estimated binary logit model (or a machine learning classifier) to obtain a propensity score of being detected, and the inverse probability weighting (IPW) (Wooldridge, 2007) could be utilized to (1) obtain an unbiased frequency of joint activities for a particular group from GLH data, (2) merge identified joint activities with GLH data and those with other data sources, and/or (3) develop a sampling scheme for other data-gathering methodologies, under a set of assumptions that need to be met to use the IPW.

Another key finding was the large gap in detection rates between iPhone and Android, which imposes a serious limitation on the usability of GLH data to study joint travel behavior, given the high market shares that the iPhone enjoy in most countries. Overcoming this issue would probably require user cooperation in GLH editing to improve data quality, at the price of a higher response burden. This gap could be partially attributed to the difference in privacy policies between iPhone and Android (Greene and Shilton, 2018) and indicates that the observed detection rates might depend to some extent on privacy policies agreed between data collectors and providers. Regarding the public acceptance of privacy-encroaching policies, recently, Lewandowsky et al. (2021) found in a study in the United Kingdom that co-location tracking can be accepted by the majority of the population when the data is used to collect contact data for infectious or potentially infectious persons during the COVID-19 pandemic. This suggests that people would have a higher willingness to provide their data when they can understand how the data is going to be utilized. However, it must be noted that privacy concerns, which might be different across cultural and socio-political contexts, might exacerbate the self-selection issue in terms of study participation.

Another major problem of relying on GLH data as a data collection tool is that the detection rates would be changing depending on the privacy policy agreement between OS firms and users. In order to collect data at the required quality level we may have to (1) identify the required level of privacy encroachment to obtain the desired accuracy levels and corresponding detection rates, (2) engage in better science communication to help people clearly understand how the data would be utilized to improve urban and transport systems, (3) confirm whether the required level of privacy encroachment can be accepted or not, and (4) have a privacy policy agreement between public bodies and citizens, separately from the ones made with OS firms.

Finally, it is important to note that the present study is a cross-section in time and space of the potential of using GLH for joint activity detection. However, it is hard to predict the potential of GLH usage in the future. On one hand, younger generations who grew up with smartphones and high-speed internet might be accustomed to higher degrees of privacy encroachment in exchange for high quality mobile experiences, which might result in higher detection accuracy levels for location services, and possible easier access to data. On the other hand, given increasing privacy concerns, we can envision, albeit with clear regional disparities, tougher privacy protection regulations being enacted, thus limiting the quantity and quality of the data that can be collected. Furthermore, changes in internal policies, and data collection methods by commercial providers can happen overnight, and it is hard to tell how these changes might affect data quality and access.

## 9. Author statement

GP conceived the study. GP, KS, YO and MC designed, and coordinated the experiments. KS processed the experiment data and conducted the detection accuracy analysis. GP and YO estimated the choice models. GP wrote the first draft, YO and MC revised the draft, and all authors approved the final manuscript.

## 10. Funding Statement

This work was supported by JSPS KAKENHI Grants Number 20H02266.

## 12. Appendix 1: GLH data example for an activity at the Koishikawa Botanical Garden in Tokyo (As of December 2020)

| "placeVisit" | | | Value |
|---|---|---|---|
| location | latitudeE7 (※) | | 357195376 |
| | longitudeE7 (※) | | 1397450839 |
| | placeId (※) | | "ChIJu8Eu6bONGGARxeXceJAa2Lg" |
| | address | | "3丁目-7-1 白山\n 文京区 東京都 112-0001\nJapan" |
| | name | | "Koishikawa Botanical Garden" |
| | sourceInfo | deviceTag | -1403052951 |
| | locationConfidence | | 87.92917 |
| duration | startTimestampMs (※) | | "1608168891941" |
| | endTimestampMs (※) | | "1608169254650" |
| placeConfidence | | | "MEDIUM_CONFIDENCE" |
| centerLatE7 | | | 357176650 |
| centerLngE7 | | | 1397464100 |
| visitConfidence | | | 89 |
| otherCandidateLocation[[1]] | latitudeE7 | | 357169991 |
| | longitudeE7 | | 1397457466 |
| | placeId | | "ChIJ7XEqobONGGARjl_w7QHFziY" |
| | locationConfidence | | 0.8224865 |
| otherCandidateLocation[[2]] | latitudeE7 | | 357177174 |
| | longitudeE7 | | 1397487859 |
| | placeId | | "ChIJPbQwSxmTGGARQm3B9Fg_j80" |
| | locationConfidence | | 0.6779116 |
| … | … | | … |
| editConfirmationStatus | | | "NOT_CONFIRMED" |
| simplifiedRawPath | points[[1]] | latE7 | 357176868 |
| | | lngE7 | 1397469228 |
| | | timestampMs | "1608168891941" |
| | | accuracyMeters | 20 |
| | points[[2]] | latE7 | 357175024 |
| | | lngE7 | 1397464561 |
| | | timestampMs | "1608169012697" |
| | | accuracyMeters | 10 |
| | … | … | … |
| **"activitySegment"** | | | **Value** |
| startLocation | latitudeE7 | | 357147506 |
| | longitudeE7 | | 1397601273 |
| | sourceInfo | deviceTag | -1403052951 |
| endLocation | latitudeE7 | | 357176868 |
| | longitudeE7 | | 1397469228 |
| | sourceInfo | deviceTag | -1403052951 |
| duration | startTimestampMs | | "1608166857669" |
| | endTimestampMs | | "1608168891941" |
| distance | | | 2201 |
| activityType | | | "WALKING" |
| confidence | | | "HIGH" |
| activities[[1]] | activityType | | "WALKING" |
| | probability | | 86.84401 |
| activities[[2]] | activityType | | "STILL" |
| | probability | | 10.627 |
| … | … | | … |
| waypointPath | waypoints[[1]] | latE7 | 357147369 |
| | | lngE7 | 1397601318 |
| | waypoints[[2]] | latE7 | 357152862 |
| | | lngE7 | 1397603454 |
| | … | … | … |
| simplifiedRawPath | points[[1]] | latE7 | 357155609 |
| | | lngE7 | 1397607269 |
| | | timestampMs | "1608166978664" |
| | | accuracyMeters | 43 |
| | points[[2]] | latE7 | 357161942 |
| | | lngE7 | 1397594299 |
| | | timestampMs | "1608167101690" |
| | | accuracyMeters | 16 |
| | … | … | … |

(※) data used in this study